\documentclass[aps,pra,reprint,superscriptaddress]{revtex4-1}
\usepackage{graphicx}
\usepackage{fullpage}
\usepackage{upgreek}
\usepackage{verbatim}
\usepackage{graphicx,epstopdf}
\usepackage{xcolor}
\usepackage{braket}
\usepackage{soul}
\usepackage{subfig}
\usepackage{amsmath}
\usepackage{siunitx}

\begin{document}

\title{Imaging individual barium atoms in solid xenon for barium tagging in nEXO}

\newcommand{\Stanford}{\affiliation{Physics Department, Stanford University, Stanford, California 94305, USA}}
\newcommand{\SLAC}{\affiliation{SLAC National Accelerator Laboratory, Menlo Park, California 94025, USA}}
\newcommand{\ORNL}{\affiliation{Oak Ridge National Laboratory, Oak Ridge, TN 37831, USA}}
\newcommand{\Yale}{\affiliation{Department of Physics, Yale University, New Haven, CT 06511, USA}}
\newcommand{\UMass}{\affiliation{Amherst Center for Fundamental Interactions and Physics Department, University of Massachusetts, Amherst, MA 01003, USA}}
\newcommand{\IME}{\affiliation{Institute of Microelectronics, Beijing, China}}
\newcommand{\Indiana}{\affiliation{Department of Physics and CEEM, Indiana University, Bloomington, IN 47405, USA}}
\newcommand{\Erlangen}{\affiliation{Erlangen Centre for Astroparticle Physics (ECAP), Friedrich-Alexander University Erlangen-N\"urnberg, Erlangen 91058, Germany}}
\newcommand{\PNL}{\affiliation{Pacific Northwest National Laboratory, Richland, WA 99352, USA}}
\newcommand{\Carleton}{\affiliation{Department of Physics, Carleton University, Ottawa, Ontario K1S 5B6, Canada}}
\newcommand{\Duke}{\affiliation{Department of Physics, Duke University, and Triangle Universities Nuclear Laboratory (TUNL), Durham, North Carolina 27708, USA}}
\newcommand{\Illinois}{\affiliation{Physics Department, University of Illinois, Urbana-Champaign, Illinois 61801, USA}}
\newcommand{\ITEP}{\affiliation{Institute for Theoretical and Experimental Physics named by A. I. Alikhanov of National Research Center “Kurchatov Institute”, Moscow 117218, Russia}}
\newcommand{\Sherbrooke}{\affiliation{Universit\'e de Sherbrooke, Sherbrooke, Qu{\'e}bec J1K 2R1, Canada}}
\newcommand{\LLNL}{\affiliation{Lawrence Livermore National Laboratory, Livermore, CA 94550, USA}}
\newcommand{\RPI}{\affiliation{Department of Physics, Applied Physics and Astronomy, Rensselaer Polytechnic Institute, Troy, NY 12180, USA}}
\newcommand{\McGill}{\affiliation{Physics Department, McGill University, Montreal, Qu{\'e}bec, Canada}}
\newcommand{\Triumph}{\affiliation{TRIUMF, Vancouver, British Columbia V6T 2A3, Canada}}
\newcommand{\IHEP}{\affiliation{Institute of High Energy Physics, Beijing, China}}
\newcommand{\Colorado}{\affiliation{Physics Department, Colorado State University, Fort Collins, Colorado 80523, USA}}
\newcommand{\BNL}{\affiliation{Brookhaven National Laboratory, Upton, New York 11973, USA}}
\newcommand{\Laurentian}{\affiliation{Department of Physics, Laurentian University, Sudbury, Ontario P3E 2C6 Canada}}
\newcommand{\SD}{\affiliation{Department of Physics, University of South Dakota, Vermillion, South Dakota 57069, USA}}
\newcommand{\Bama}{\affiliation{Department of Physics and Astronomy, University of Alabama, Tuscaloosa, AL 35487, USA}}
\newcommand{\Drexel}{\affiliation{Department of Physics, Drexel University, Philadelphia, Pennsylvania 19104, USA}}
\newcommand{\Stony}{\affiliation{Department of Physics and Astronomy, Stony Brook University, SUNY, Stony Brook, New York 11794, USA}}
\newcommand{\IBS}{\affiliation{IBS Center for Underground Physics, Daejeon 34047, Korea}}
\newcommand{\LHEP}{\affiliation{LHEP, Albert Einstein Center, University of Bern, Bern, Switzerland}}
\newcommand{\CalTech}{\affiliation{Kellogg Lab, Caltech, Pasadena, California 91125, USA}}
\newcommand{\UNCW}{\affiliation{Department of Physics and Physical Oceanography, UNC Wilmington, Wilmington, NC 28403, USA}}

\author{C.~Chambers} \Colorado
\author{T.~Walton} \altaffiliation{now at Prism Computational Sciences
Madison, WI}\Colorado
\author{D.~Fairbank} \Colorado
\author{A.~Craycraft} \Colorado
\author{D.R.~Yahne} \Colorado
\author{J.~Todd} \Colorado
\author{A.~Iverson} \Colorado
\author{W.~Fairbank} \Colorado

\author{A.~Alamre}\ITEP
\author{J.B.~Albert}\Indiana
% ...

\collaboration{nEXO Collaboration}

\date{\today}

\begin{abstract}
\textbf{The search for neutrinoless double beta decay probes the fundamental properties of neutrinos, including whether or not the neutrino and antineutrino are distinct.  Double beta detectors are large and expensive, so background reduction is essential for extracting the highest sensitivity. The identification, or ``tagging'', of the \textsuperscript{136}Ba daughter atom from double beta decay of \textsuperscript{136}Xe provides a technique for eliminating backgrounds in the nEXO neutrinoless double beta decay experiment. The tagging scheme studied in this work utilizes a cryogenic probe to trap the barium atom in solid xenon, where the barium atom is tagged via fluorescence imaging in the solid xenon matrix. Here we demonstrate imaging and counting of individual atoms of barium in solid xenon by scanning a focused laser across a solid xenon matrix deposited on a sapphire window. When the laser sits on an individual atom, the fluorescence persists for $\sim$30~s before dropping abruptly to the background level, a clear confirmation of one-atom imaging. No barium fluorescence persists following evaporation of a barium deposit to a limit of $\leq$0.16\%. This is the first time that single atoms have been imaged in solid noble element. It establishes the basic principle of a barium tagging technique for nEXO.}
\end{abstract}


\maketitle 

The search for neutrinoless double beta decay ($0\nu\beta\beta$) is an important probe into the nature of neutrinos. Observation would imply that neutrinos are Majorana particles, would demonstrate violation of lepton number conservation, and could help determine the absolute neutrino mass\textsuperscript{1}. EXO-200 is searching for $0\nu\beta\beta$ in \textsuperscript{136}Xe with 110~kg of active liquid Xe (LXe) enriched to 80.6\si{\percent} \textsuperscript{136}Xe in a time projection chamber (TPC).  Two-neutrino double beta decay ($2\nu\beta\beta$) of \textsuperscript{136}Xe has been observed in EXO-200, and its half-life is measured at $T^{2\nu\beta\beta}_{1/2} = 2.165 \pm 0.016$(stat)$ \pm 0.059$(sys)$\times 10^{21}$~yr \cite{EXO200TwoNuLong}. The most recent EXO-200 $0\nu\beta\beta$ search sets a limit on the half-life at $T^{0\nu\beta\beta}_{1/2}$~\textgreater~1.8$\times$10$^{25}$~yr (90\si{\percent} CL), which corresponds to an effective Majorana neutrino mass of $\braket{m_{\nu_{e}}}$~\textless~147-398~\si{\milli\electronvolt}, depending on nuclear matrix element calculations \cite{EXO200ZeroNuPRL}. 

A \textsuperscript{136}Xe TPC provides a unique opportunity to tag the daughter \textsuperscript{136}Ba at the site of a double beta decay event. The implementation of this barium tagging would improve $0\nu\beta\beta$ sensitivity by effectively eliminating all backgrounds except $2\nu\beta\beta$ \cite{Moe1991}. Barium tagging is being investigated for a future upgrade of the next-generation LXe experiment, nEXO, a 5~tonne enriched Xe experiment recently described in \cite{pCDR,sensitivity}. Initial results have been reported for research on methods of barium tagging in LXe \cite{Mong2015,Twelker2014}, and also in a Xe gas TPC \cite{Brunner2015}. The NEXT collaboration has recently reported images of single Ba$^{++}$ in fluorescent dye molecules from a dilute deposit of barium perchlorate salt solution \cite{NEXTPRL}.

This paper presents a major step towards realization of barium tagging in solid Xe (SXe) for nEXO ~\cite{Mong2015}. In this method, a cryogenic probe would be moved to the position of the $0\nu\beta\beta$ candidate event in LXe, and the daughter atom or ion would be captured in a small amount of SXe on a sapphire window at the end of the probe~\cite{Wamba,Rowson}. It would then be detected by its laser-induced fluorescence in the SXe. It is expected that a Ba\textsuperscript{++} ion will convert to Ba\textsuperscript{+} in LXe, as the LXe conduction band gap is less than the ionization potential for Ba\textsuperscript{+}~\cite{Moe1991}.  Neutralization to barium may also occur in the charge cloud following a $\beta\beta$ event. A study of $^{214}$Bi daughters of $^{214}$Pb $\beta$-decay in EXO-200 has reported that 76(6)\si{\percent} of these daughters are ionized, with negligible subsequent neutralization after many minutes~\cite{alphaion}. Thus, a large fraction of $^{136}$Ba $0\nu\beta\beta$ daughters is expected to be in the singly ionized state in LXe. Whether or not the $^{136}$Ba will remain ionized in SXe on a cold probe is not yet known.

\begin{figure}
\includegraphics[width=0.45\textwidth]{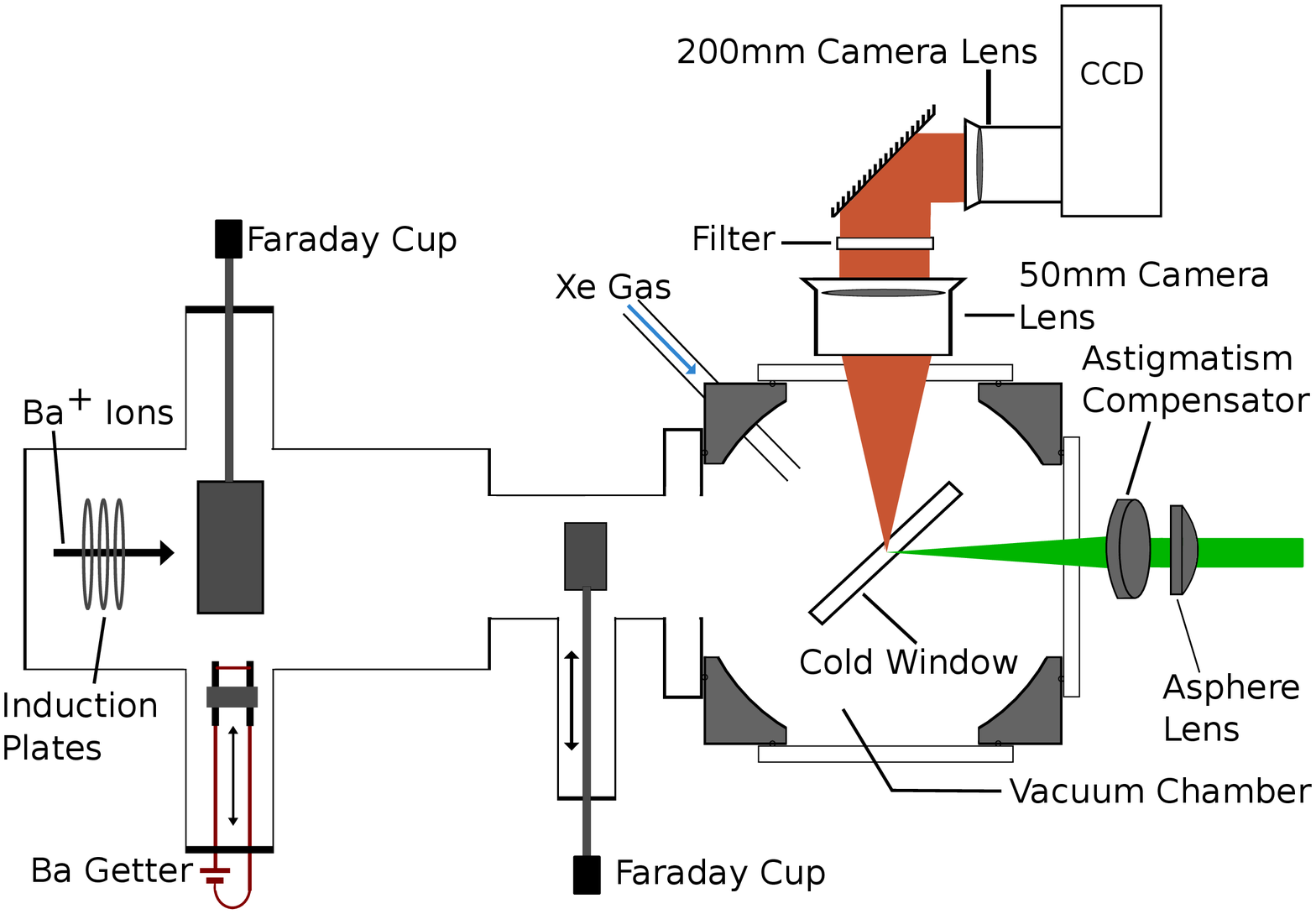}
\caption{\textbf{Experimental setup for Ba/Ba\textsuperscript{+} imaging in solid Xe.} The two sources of barium are shown at the left of the figure, the BaAl$_4$ getter and the pulsed Ba$^{+}$ ion beam. The ion pulses are measured with a set of induction plates and two Faraday cups, which ensure the alignment of the ion beam.Xe gas is condensed on a 0.5~mm thick sapphire window tilted at 45\textsuperscript{$\circ$} mounted to a cryostat cold finger. The Xe gas enters the cryostat via a tube pointed at the upper window surface. Pulses of Ba$^{+}$ ions are incident during th Xe deposit. The neutralized barium atoms are excited by a dye laser focused by a 7.9~cm focal length lens, with an optical flat used to compensate for the astigmatism introduced by the tilt of the sapphire window. The lens is mounted on piezo-electric translation stages, allowing the laser to be precisely moved across the sample. The fluorescence is observed from above, collimated by a 50~mm camera lens and focused onto a liquid nitrogen cooled CCD camera by a 200~mm camera lens. The fluorescence is filtered to pick out the emission range of interest.}
\label{fig:apparatus}
\end{figure}

Significant progress on understanding the spectroscopy of Ba in SXe has been made \cite{Mong2015,McCaffrey2016,McCaffrey2018}. Through theoretical modeling, the strongest fluorescence peaks at 577 and 591~\si{\nano\meter} are identified as barium atoms in 5-atom and 4-atom vacancy sites in the SXe matrix~\cite{McCaffrey2018}. These two fluorescence peaks bleach fairly rapidly at high laser intensity, e.g. using a focused laser beam~\cite{Mong2015}. Obtaining large numbers of photons from single barium atoms in these matrix sites would benefit from a method to overcome bleaching, e.g., with repumping lasers. The barium emission peak at 619~\si{\nano\meter}, although not reported in some deposits made at 10~\si{\kelvin}~\cite{McCaffrey2016,McCaffrey2018}, is more prominent in annealed deposits or deposits made at higher temperature and observed at 10~\si{\kelvin}~\cite{Mong2015,McCaffrey-pc}. The 619~\si{\nano\meter} peak is attributed in this work to barium atoms in a single vacancy (SV) site. This peak exhibits less bleaching than the 577 and 591~\si{\nano\meter} emission peaks and thus is more amenable to single barium atom imaging. The apparatus for depositing and observing Ba/Ba\textsuperscript{+} deposits in SXe is described in \cite{Mong2015}. Important components are shown in Fig. \ref{fig:apparatus}.

In this work, imaging of single barium atoms in SXe via the 619~\si{\nano\meter} fluorescence peak is reported. This is the first time that single atoms have been imaged in a solid noble element matrix. Images of single DBATT dye molecules in solid krypton and xenon~\cite{Sepiol} as well as Mg-TAP molecules in solid xenon~\cite{Starukhin} have been obtained previously.

\section{Results}
\label{sec:results}

\subsection{Fixed Laser Images}
\label{sec:fixedimages}

\begin{figure}
\includegraphics[width=0.48\textwidth]{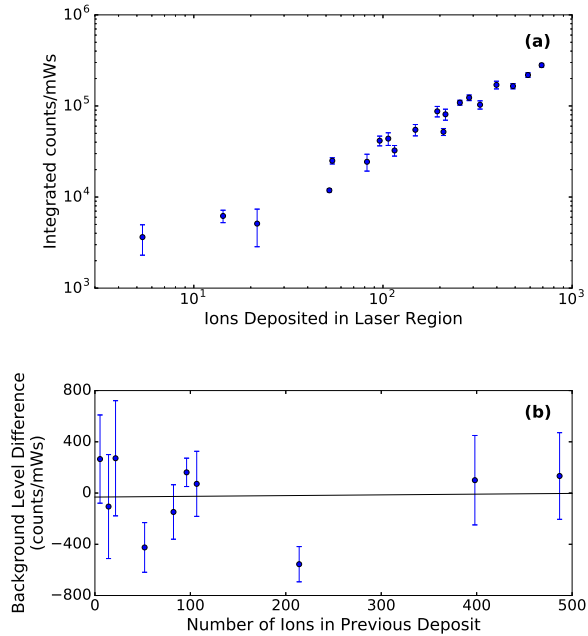}
\caption{\textbf{Barium atom fluorescence.} a) 619~nm Ba fluorescence vs. number of Ba\textsuperscript{+} ions deposited. The detected CCD counts have been scaled by the laser exposure in mW$\times$s (counts/mWs). The slope of a linear fit to the data is 379~$\pm$~10~counts/mWs per ion, with a log-log slope of 1.04~$\pm$~0.05. b) The difference in background level from SXe-only from before and after a barium deposit is plotted with respect to the number of barium ions in the deposit. The weighted least-squares fit (black line) has a slope of 0.06~$\pm$~0.41. In both panels, the data points are the averages and the error bars are the standard deviations of the four observation locations.}
\label{fig:ctsVsIons}
\end{figure}

The barium fluorescence in a given deposit is determined by summing the counts in a 4$\times$4 pixel area enclosing the laser spot and subtracting the SXe-only background. The background is measured by averaging the summed CCD counts in the focused laser region from the prior and following SXe-only deposits for each Ba\textsuperscript{+} deposit. A typical background level is $\sim$1000~\si[per-mode=symbol]{counts\per\milli\watt\second}. The observed barium counts per mWs of laser exposure vs. Ba\textsuperscript{+} ions deposited in the laser region are plotted in Fig. \ref{fig:ctsVsIons}(a). Since the neutralization fraction is not known, the number of barium atoms in the laser beam is less than or equal to the number of ions deposited. Each point represents a separate Ba\textsuperscript{+} deposit with the signal averaged over 4 laser positions separated by 20~\si{\micro\meter} on the deposit. The error bar is the standard deviation of the four measurements. The observed signal is linear with a log-log slope of \num[separate-uncertainty]{1.04\pm0.05}. The slope of the linear fit is \num[separate-uncertainty]{379\pm10}~\si[per-mode=symbol]{counts\per\milli\watt\second} per ion. Including the uncertainty in number of ions deposited, the observed fluorescence per ion is 379$^{+39}_{-76}$~\si[per-mode=symbol]{counts\per\milli\watt\second} per ion. For these measurements, $\sim$40~\si{\micro\watt} of focused 572~\si{\nano\meter} laser excitation was used with $\sim$3~\si{\second} of laser exposure.

For a practical barium tagging application in nEXO, it is crucial for each barium tag to be independent of any previous tagging measurements or residual barium on the cryogenic probe to ensure that the barium daughter is correlated to the decay candidate being investigated. To obtain a quantified limit of this ``erasure'' property, the difference in the background level of the last SXe-only deposit before and the SXe-only immediately after a barium deposit in Fig.~\ref{fig:ctsVsIons}(a) is plotted in Fig.~\ref{fig:ctsVsIons}(b).  The data are averaged over four laser positions for each deposit, and the standard deviation is shown as an error bar. The differences are consistent with zero, and the slope of a linear fit, \num[separate-uncertainty]{0.06\pm.41}~\si[per-mode=symbol]{counts\per\milli\watt\second} per ion is a measure of the barium signal remaining after evaporation. The ratio of the maximum value of the SXe-only difference fit to the minimum barium signal response from the slope of Fig.~\ref{fig:ctsVsIons}(a) gives a limit of $\leq$~0.16\% on residual barium fluorescence after evaporation. 

\subsection{Scanned Images}
\label{sec:scanning}

\begin{figure*}
\includegraphics[width=1.0\textwidth]{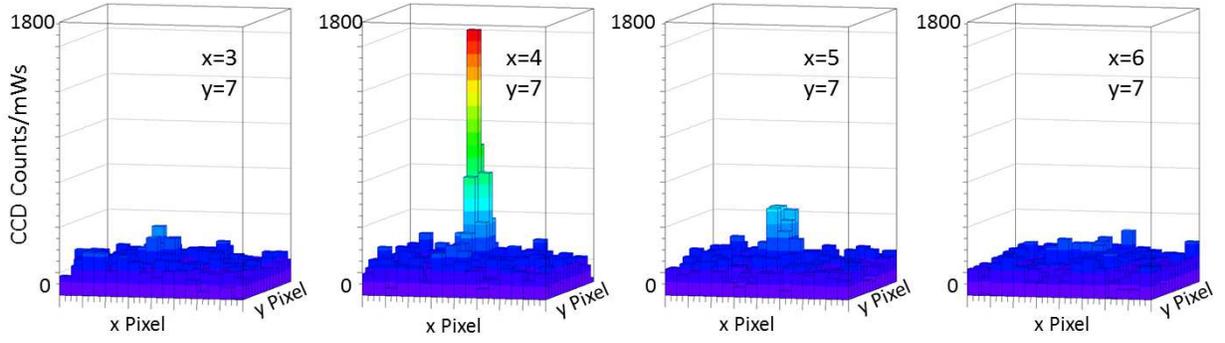}
\caption{\textbf{CCD images of successive steps of a raster scan of a barium in solid Xe deposit.} As the laser passes over a barium atom from left to right, the signal abruptly increases when the laser is positioned at the location of the barium atom. The (x,y) coordinates refer to the laser displacement in 4~$\mu$m scan steps. This corresponds to 4~$\mu$m and 5.7~$\mu$m steps, respectively, on the sample  The detected CCD counts have been scaled by the laser exposure in mW$\times$s (counts/mWs).}
\label{fig:scan_steps}
\end{figure*}

By further decreasing the density of barium ions deposited, single barium atoms can be spatially resolved.  To image these atoms, the laser is rastered across the deposit. A typical scan consists of laser displacements in a square grid of 12$\times$12 steps with 3~\si{\second} of laser exposure and a spacing of 4~\si{\micro\meter}.  The raw CCD images of four successive steps of a laser scan are shown in Fig.~\ref{fig:scan_steps}. As the laser passes over the barium atom, a strong 619~\si{\nano\meter} signal appears. When the laser moves off of the barium atom, the observed counts return to background level. 

\begin{figure*}
\includegraphics[width=1.0\textwidth]{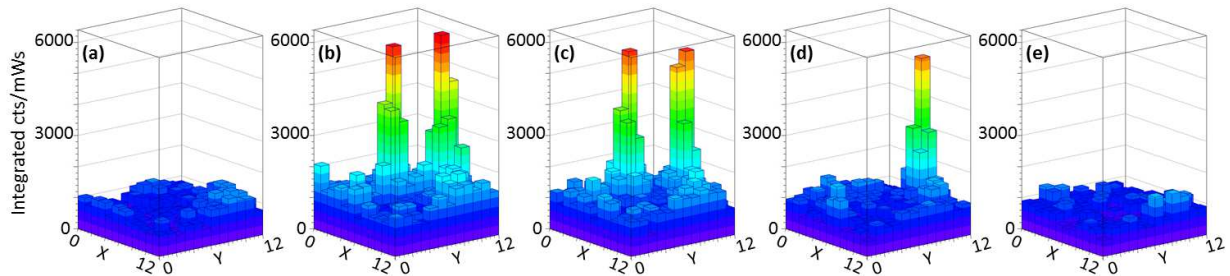}
\caption{\textbf{Composite images of a sequence of laser scans.} The laser displacement step size is 4~$\mu$m, with a 12$\times$12 grid. The integrated CCD counts have been scaled by the laser exposure in mW$\times$s (counts/mWs). First, a scan of a SXe-only deposit is done (a) and evaporated. Another deposit is done with barium in SXe and scanned twice, (b) and (c). A third scan of the barium in SXe deposit is then done (d) after observing the left barium atom peak for 150~s, during which observation it disappears. The barium in SXe sample is then evaporated and another SXe-only deposit made and scanned (e).}
\label{fig:scan}
\end{figure*}

A scan image of this region is generated by integrating the CCD counts in a 4$\times$4 pixel area encompassing the laser region in each frame.  These integrals are divided by the laser exposure, and arranged according to the position of each spot in the grid. Five such scan images, taken from a typical experiment, are shown in Fig.~\ref{fig:scan}. Data from the four frames shown in Fig~\ref{fig:scan_steps} contribute to four points in line y=7 in Fig.~\ref{fig:scan}(b). In the first SXe-only deposit, no fluorescence peaks are found in the scan (Fig.~\ref{fig:scan}(a)). After this, the deposit is evaporated by heating the sapphire window to 100~\si{\kelvin}. Then, a new SXe deposit with 12 pulses of the Ba$^{+}$ ion source, corresponding to 48$^{+5}_{-10}$~Ba$^{+}$ ions in the full scan area is produced. Two large peaks of roughly equal size in an otherwise low-background area are observed in a scan of this deposit (Fig.~\ref{fig:scan}(b)). This distinguishes each peak as the location of a single resonant barium atom. This scan is then repeated, and the barium peaks persist, as seen in Fig.~\ref{fig:scan}(c). 

\begin{figure}
\includegraphics[width=.5\textwidth]{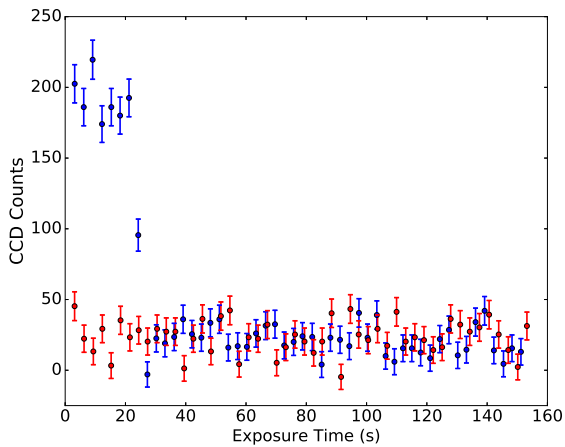}
\caption{\textbf{Time evolution of the fluorescence signal from a single barium peak.} The signal from the location of an atom is shown in blue, and is compared to the signal background from a SXe-only run shown in red. The error bars are the uncertainties from readout noise and photon statistics added in quadrature. In these runs, the average laser exposure per frame is 5.5$\times$10$^{-2}$~mWs.}
\label{fig:singlespot}
\end{figure}

At this point, the laser is moved to the location (x,y)=(4.75,6.00), and many 3~\si{\second} exposures are taken. Although not intended, this position is not the same as the peak frame (4,7) but is close to the position of frame (5,6). The time dependence of the integrated signal from a 3$\times$3 pixel area of the single barium peak is shown in Fig.~\ref{fig:singlespot}. The average signal level of 192~counts agrees with the single barium values at this position in the scans, 184~and~169~counts, respectively, taking into account 3$\times$3 rather than 4$\times$4 pixel integration. This signal persists for $\sim$30s of laser exposure, including the prior two scans, before abruptly dropping to the average background level of 21~counts. This discrete turn-off of the fluorescence signal is a hallmark of a single atom.

About 3300 photons (0.5 CCD \si[per-mode=symbol]{counts\per photon}) are detected from this atom. This corresponds to around 1.4$\times$10$^{6}$ photons absorbed and emitted by one atom. For comparison, the standard deviation of the SXe-only background at one laser position in Fig.~\ref{fig:singlespot} for a 30~\si{\second} integration is 47 detected photons. Thus, the single barium atom fluorescence signal is 70~$\sigma$ above background fluctuations at one laser position. For different laser positions in the SXe-only scans, Fig.~\ref{fig:scan} (a) and (e), the background fluctuation~$\sigma$ is 30 detected photons in a 3~\si{\second} integration. Comparing this to 330 barium atom fluorescence photons in 3\si{\second}, the barium peaks in a composite image, such as those in Fig.~\ref{fig:scan} (b-d), are 11~$\sigma$ above background.

Following this fixed laser position run, a third scan is done, and the barium peak on the left has disappeared, as expected. The right barium peak on the edge of the scan persists, as seen in Fig.\ref{fig:scan} (d). After evaporating this deposit, a new SXe-only deposit is made. As in the previous SXe-only deposit, no peaks are observed in this scan (Fig.\ref{fig:scan} (e)). This illustrates the lack of any ``history~effect" due to previous barium deposits and the absence of any signal from possible barium contamination on the sapphire window.

\section{Discussion}

\begin{figure}
\includegraphics[width=0.48\textwidth]{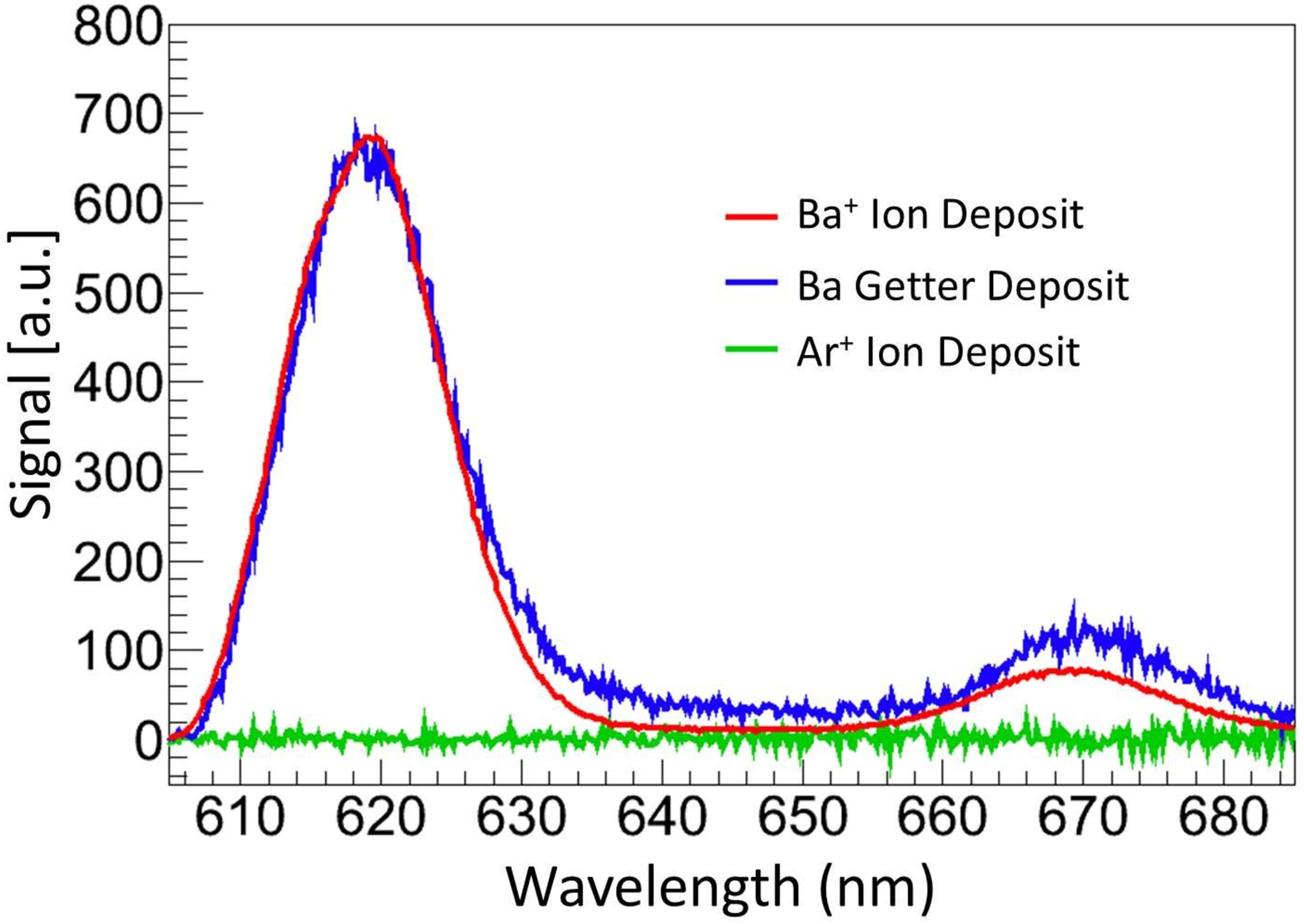}
\caption{\textbf{Spectra of deposits from three different sources in SXe.} The barium getter and barium ion beams produce the same fluorescence peaks, which are not produced by matrix damage from an argon ion beam. The barium getter deposit spectrum is scaled since the deposited barium density cannot be determined.}
\label{fig:ion_getter_ar}
\end{figure}

The barium sample in this work is deposited as Ba\textsuperscript{+} ions. The fluorescence lines of Ba\textsuperscript{+} in SXe are expected in the blue-green region rather than the yellow-red region. Thus a spectroscopic assignment of the 619~\si{\nano\meter} emission line to barium rather then Ba\textsuperscript{+} is favored. To further test this, a spectrum of a neutral barium deposit made with the barium getter source is compared in Fig.~\ref{fig:ion_getter_ar} to a spectrum of a Ba$^{+}$ deposit. Identical spectra are observed using the two sources under similar conditions. This confirms the 619~\si{\nano\meter} emission line as associated with barium atoms resulting from neutralization of the incident Ba$^{+}$ ions. In addition, no fluorescence peak is observed from deposits of Ar$^{+}$ ions in SXe at 2000~\si{\electronvolt} under similar conditions. This rules out matrix damage as the source of the 619~\si{\nano\meter} peak. 

A reasonable assignment of the 619 nm emission is to barium atoms in a single vacancy (SV) site. The previously studied case of Na$^{+}$ in solid argon (SAr) matrix is a good analogy. A higher fraction of Na atoms was found in SV sites with Na$^{+}$ ions incident from a laser ablation or ion beam source than with Na atoms from a thermal atom source~\cite{Tam-Fajardo,Silverman-Fajardo,Vaskonen,Ahokas}. In the model presented in~\cite{Silverman-Fajardo}, Na$^{+}$ ions preferentially form in SV sites in SAr due to tighter ion binding. Subsequently, some ions are neutralized, resulting in Na atoms in cramped SV sites. Similarly, a barium ion with Ba$^{+}$Xe equilibrium radius of 3.619~\si{\angstrom}~\cite{Buchachenko} should prefer an SV site of 4.39~\si{\angstrom} radius in SXe at 50~\si{\kelvin}~\cite{Granfors} rather than a larger multivacancy site. After neutralization from Ba$^{+}$ to Ba, the barium atom with BaXe equilibrium radius of 5.5~\si{\angstrom}~\cite{Buchachenko,McCaffrey2016,McCaffrey2018} is cramped in the SV site, but creation of additional vacancies may not be energetically favorable. In the following paragraphs, alternate barium molecule interpretations are considered, and found to be inconsistent with observations.

Assignment to Ba$_2$ is unlikely as a linear relationship of signal vs. ions deposited is observed in Fig.~\ref{fig:ctsVsIons}(a), rather than quadratic. At the low ion density of Fig.~\ref{fig:scan}, implanted Ba$^{+}$ ions are separated, on average, by 8~\si{\micro\meter}. Since the Ba\textsuperscript{+} ions are implanted 7~$\pm$~5~\si{\nano\meter} \cite{SRIM} below the surface of the SXe, mobility of the neutralized barium atoms should be limited.

Reactions of deposited Ba with residual gas impurities, such as water and oxygen to form molecules should also be considered. A limit of \textless~100~\si{ppm} residual gas molecules in the SXe matrix is established through fringe rate and partial pressure comparisons. The most likely species to form are BaOH and BaO, through the reactions

\begin{equation}
\begin{split}
\mathrm{Ba + H_{2}O}&\rightarrow \mathrm{BaOH + H} \\
\mathrm{Ba + O_{2}}&\rightarrow \mathrm{BaO + O} \\ 
\end{split}
\end{equation}
 
Both processes are endothermic \cite{Murad1981}, and thus energetically unfavored. BaOH vapor has known emission bands in the green at 487~\si{\nano\meter} and 512~\si{\nano\meter}, and in the IR at 712-758~\si{\nano\meter} and 783-839~\si{\nano\meter} \cite{BaOHlines}. These do not agree with the observed 619~\si{\nano\meter} emission. At low temperature, the upper state of BaO should be primarily the A$^{1}\Sigma^{+}$($\nu'$) state. This state has a multi-line spectrum in vacuum with calculated values from 455 - 865~\si{\nano\meter} \cite{BaOlines}. The $\nu'$=0 to $\nu''$=2 and 4 transitions are close to our observed peaks at 619~nm and 670~\si{\nano\meter}, but adjacent transitions are missing. Furthermore, the excitation spectra of these peaks are different, contradicting a common $\nu'$=0 upper state for the emission. The A$^{1}\Sigma^{+}$ state lifetime of 356~\si{\nano\second} in vacuum~\cite{Johnson1972}, is discrepant with the 7~\si{\nano\second} decay lifetime of the observed 619~\si{\nano\meter} emission line presented in this work. Thus, this line is not due to~BaO.

\section{Conclusions}

The imaging and counting of individual barium atoms with high definition in solid xenon by scanning a focused laser has been demonstrated. The 619~\si{\nano\meter} emission peak observed in deposits of Ba$^{+}$ and Ba in solid xenon is attributed to neutral barium atoms in a single-vacancy substitution site. The observation of two isolated peaks in a scan with roughly equal size fluorescence in an otherwise low-background area clearly distinguishes each peak as the location of an individual resonant barium atom.  With prolonged laser exposure, the sudden, discrete turn-off of the fluorescence, with signal dropping to background level, is strong confirming evidence that the fluorescence is from one atom only.  This is the first imaging of individual atoms in a solid rare element. Additional unique features of this work are very low background obtained by pre-bleaching the substrate and a demonstrated limit of $\leq$0.16\% for ``erasure'' of barium signal by deposit evaporation.  Successful counting of individual barium atoms in solid xenon is a significant step toward barium tagging in the nEXO neutrinoless double beta decay experiment.

\section*{Acknowledgements}

The authors are grateful to Picoquant for the loan of the time resolved photon counting equipment. Valuable discussions with J. G. McCaffrey, B. Gervais and A. van Orden are appreciated. This material is based upon work supported by the National Science Foundation under Grant Number PHY-1649324 and the U.S. Department of Energy, Office of Science, Office of High Energy Physics under Award Number DE-FG02-03ER41255.

\clearpage

\section*{Methods}

\textbf{Ba/SXe Sample Deposition.} The source of barium is an ion beam at 2~keV energy, filtered to select Ba\textsuperscript{+} with an E$\times$B velocity filter. A set of pulsing plates produces $\sim$1~$\mu$s ion bunches for depositing small numbers of ions. The spectra of Ba\textsuperscript{+} ion deposits in the SXe matrix exhibit peaks known to be due to neutral Ba atoms [7]. Thus some percentage of the ions neutralize in the matrix, although the fraction has not yet been determined. An alternative source of neutral Ba is a BaAl$_{4}$ getter wire which can be moved into the beam path and heated to emit barium atoms toward the sample. However, it is challenging to achieve low barium flux with this source and to calibrate it.

Deposits are made on a cold sapphire window tilted at 45\textsuperscript{$\circ$} with respect to the Ba\textsuperscript{+} beam. To create a sample, Xe gas is directed toward the window by opening a leak valve. The Xe gas freezes onto the window and forms a SXe matrix with a thickness of around a micron. This is initiated a few seconds prior to the barium deposit, continues during the barium deposit, and is turned off a few seconds after the barium deposit by closing the valve. Thus, the barium atoms are located sparsely in a thin layer at about half the depth of the SXe.

In this work, deposition is done with the sapphire window at a temperature of $\sim$50~K. This reduces hydrogen, nitrogen and oxygen content in the matrix, as these residual gases condense below 50~K in vacuum [7]. The window is then cooled to 11~K for observation. Xe is deposited at a rate of around 60~nm/s. An experiment cycle consists of a deposit at 50~K, a fluorescence observation at 11~K, and then evaporation of the deposit by heating the window to 100~K. Many deposits are made in a day with varying numbers of ions deposited, as well as periodic SXe-only deposits to establish the background.

\textbf{Measurement of Ba ion density.} The area density of deposited Ba\textsuperscript{+} ions cannot be measured directly because the window is an insulator. During the deposit, only the induction signal of the pulse of ions in transit through a circular induction plate is recorded. To estimate the deposited ion density, the pulsed ion beam is sampled before and after deposits by two Faraday cups located 17.5~cm and 4.5~cm before the window. The factor for conversion of these signals to ion density on the window is measured in a separate calibration procedure in which the window is replaced by a third Faraday cup at the window position, and the alignment and magnitude of the three cup signals are compared under conditions similar to that of actual deposits. The uncertainty in the Faraday cup measurements is estimated as $\pm$-10\% due to secondary electron effects, measured by biasing the Faraday cup electrodes. An additional +0\% to -10\% uncertainty in the ion density deposited on the observation window is included for possible 1~mm misalignment of the approximately Gaussian ion beam, with a full width at half maximum of 5.1~mm, relative to the excitation laser. Conservatively adding these two uncertainties linearly, the total uncertainty in barium ion density at the window is +10\% to -20\%.

The number of Ba\textsuperscript{+} ions deposited within the 1/e radius of the laser beam gives a rough upper limit to the number of barium atoms responsible for the observed signal with a fixed laser beam. For typical ion bunch densities of 0.1-1~fC/mm\textsuperscript{2} and focused laser 1/e radii of w$_{0x}$ = 3.2~$\mu$m and w$_{0y}$ = 3.8~$\mu$m, this results in about 0.001-0.01 Ba\textsuperscript{+} ions/pulse in the 1/e intensity laser region.

\textbf{Excitation laser system.} The excitation laser, a Coherent 599 cw dye laser with Rhodamine~6G dye, pumped by the 532~nm line of a Coherent Verdi-V8 laser, enters from the back side of the window. To position the laser beam with sub-$\nu$m precision, two computer controlled piezoelectric translation stages are used to move the laser focusing lens. Barium fluorescence light is collected and collimated by a 50~mm Nikon camera lens. A filter with a sharp-edged band-pass of 610-630~nm passes just the 619~nm fluorescence peak. A 200~mm Nikon camera lens then focuses the light onto a liquid nitrogen cooled CCD [31], resulting in an image of $4\times$ magnification. Each of the 20$\times$20~$\mu$m pixels of the CCD represents approximately a 5$\times$7~$\mu$m area on the SXe sample, which is at tilted at 45\textsuperscript{$\circ$} in the y-direction.

For a given laser intensity, the smallest focus possible is desired for optimal signal-to-background ratio from single atoms. To achieve this, an aspherical lens of 7.9~cm focal length [32] is used to minimize spherical aberration, and a fused silica optical flat of 1~cm thickness is placed at 9\textsuperscript{$\circ$} after the lens in order to compensate for astigmatism caused by the tilted sapphire window. 

Vibrations of the sapphire window relative to the excitation laser increase the area of laser exposure, and reduce the average intensity seen by a single barium atom.  The main source of vibration is the cryostat He compressor cycle, which pulses with a frequency of about 2.25~Hz. To limit the laser exposure to a segment of the cryostat cycle with minimal vibration, a shutter is placed in the laser path and synchronized with the signal from an accelerometer on the outside of the cryostat. This laser gating has 45\% duty cycle.
 
\textbf{Backgrounds for 619~nm emission.} A typical CCD image recorded with a focused laser beam at 570~nm is shown in Extended Data Fig.~1. A strong signal from the barium deposit in SXe on the front surface of the window is visible. A Gaussian fit to the image gives a 1/e$^{2}$ radius of 10.4~$\mu$m, which is larger than the average laser beam radius of w~=~3.5~$\mu$m. Aberrations and vibrations in the collection optics and imperfections in the surface of the SXe layer contribute to blurring of the image. A weak background emission from the opposite window surface is also visible.

Very low concentrations of Cr\textsuperscript{3+} in the sapphire bulk (sub-ppb level) produce a sharp fluorescence peak at 693~nm, with a broad tail extending to the 610-630~nm region passed by the band-pass filter.  This results in the faintly visible line through the window in the CCD image. Commercially available c-plane quality sapphire from Meller Optics and Rubicon Technologies has been found to have sufficiently low Cr\textsuperscript{3+} concentrations for detecting single barium atoms.

The background emission from the front sapphire surface beneath the SXe layer is the main challenge for single barium imaging. It has been found that this background can be reduced significantly and semi-permanently by photo-bleaching.  A variety of wavelengths have been used effectively for bleaching, including 514.5~nm, 532~nm, 570-572~nm, and 580.5~nm. A typical bleaching procedure consists of a repeating raster scan of an 80~mW 532~nm laser, focused to around w$_{0}$ = 10~$\mu$m, in a 14$\times$14~position grid, with 8~$\mu$m grid spacing and 20~$\mu$m at each position per scan. This is done with the sapphire window at 100~K. As seen in a subsequent imaging scan in Extended Data Fig.~2, this reduces the surface background $\sim$30$\times$ over a region of about 90$\times$90~$\mu$m. This area is large enough to accommodate both the fixed laser and scanned images presented in Sec.~\ref{sec:scanning} in the main article text.

\textbf{Time resolved photon counting.} The decay lifetimes of both the sapphire surface background and the average signal from many barium atoms were investigated using a 561~nm pulsed laser with 100~ps pulse length as the excitation source, and a single photon avalanche photodiode (SPAD) as the detector. The time between the laser pulse and the arrival of a photon at the SPAD was measured by a fast counter, and a histogram of photon arrival times was recorded [33]. The decay histograms for a SXe-only (green) and Ba in SXe (blue) deposit, are shown in Extended Data Fig.~3. By subtracting the SXe-only histogram from the barium histogram, the decay lifetime for the 619~nm emission of barium was isolated and measured to be 7.0~$\pm$~0.3~ns. The SXe-only background decay is comprised of more than one decay constant, but is nonetheless significantly shorter than the barium decay lifetime. By time gating the CCD or SPAD, the signal to background ratio can be increased by a factor of 2 with a loss of 50\% of the barium signal. In the future, a sapphire window will be mounted at the end of a cryoprobe~\cite{Mong2015}, likely to be made of some type of metal. Background emission from the metal tubing may be a concern for barium atom detection. The observed lifetime of emission at 619~nm from a stainless steel cryoprobe tube, shown in Extended Data Fig.~3~(red), is $\sim$1.5~ns. Thus, background from the stainless steel parts of a cryoprobe could be reduced by time gating with minimal barium signal loss.\newline 

\vspace{5mm}

\noindent
[31] Roper Scientific SPEC-10 LN-Cooled
 
\hspace{3.2mm}Spectroscopy Detector.

\noindent
[32] Thorlabs part ASL10142-A.

\noindent
[33] Picoquant PicoHarp 300, PDL800D laser and 

\hspace{3.2mm}PDM detector.

\clearpage

\begin{figure*}
\includegraphics[width=.4\paperwidth]{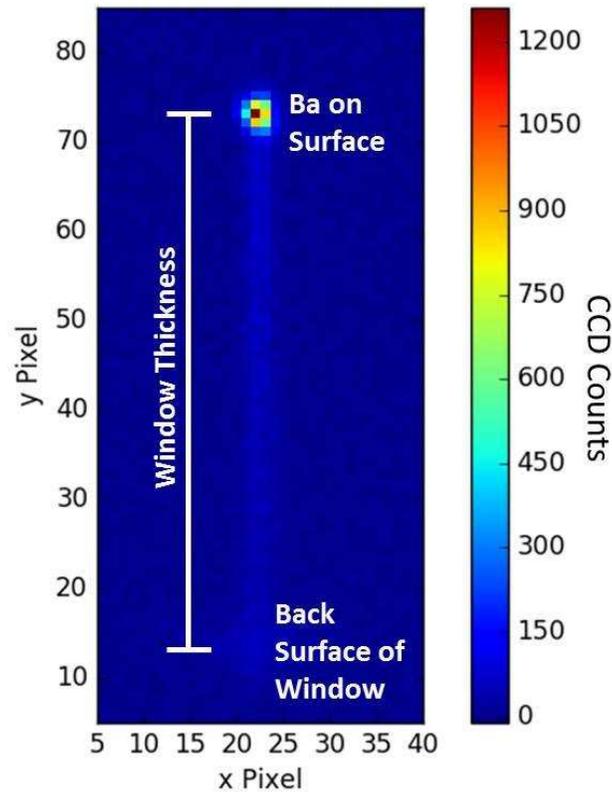}
\caption*{\textbf{Extended Data Fig. 1: Example CCD image of a Ba\textsuperscript{+} deposit in SXe.} The barium atoms are excited by a focused 570~nm laser, using a 620~nm fluorescence band-pass filter. The bright spot at the top of the image is the front surface of the window where the Ba ions are deposited. The broad spot at the bottom of the image is the surface fluorescence of back surface of the window. This spot is broadened due to the laser focus as well as the collection optics being optimized for the front surface. The faint line between the surfaces is the faint fluorescence of Cr$^{3+}$ impurities in the bulk of the sapphire that extend into the filter region.}
\end{figure*}

\begin{figure*}
\includegraphics[width=.65\paperwidth]{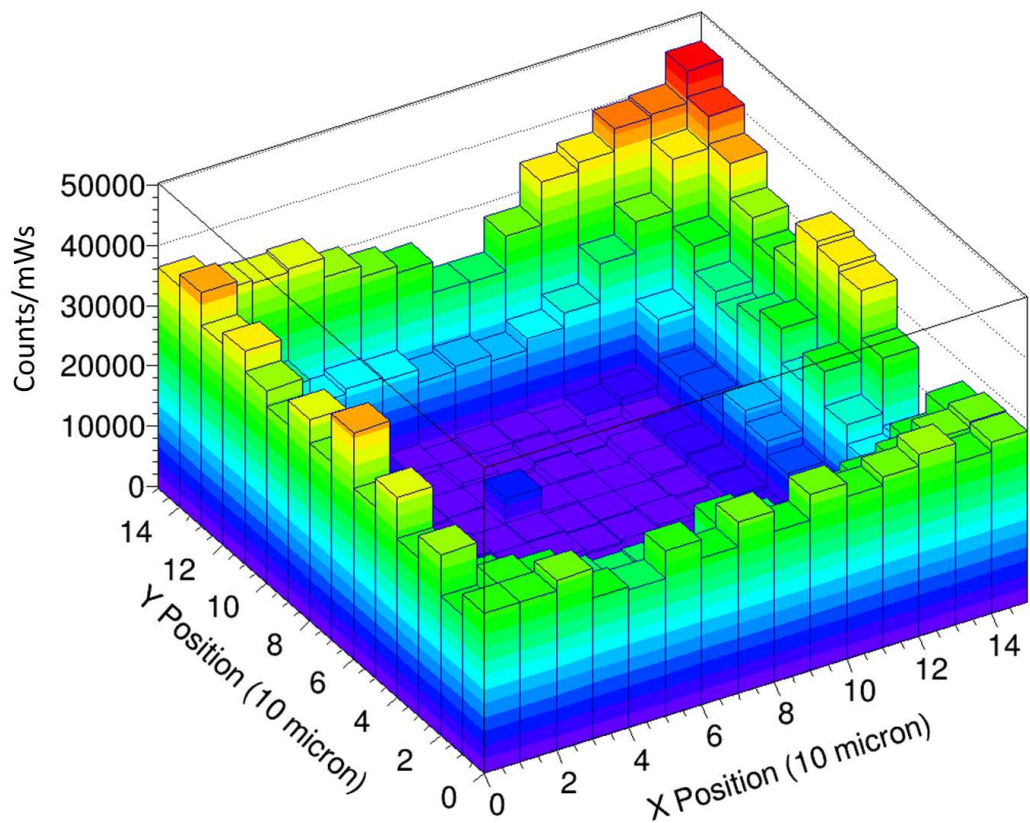}
\caption*{\textbf{Extended Data Figure 2: Scan image of background emission after bleaching.} A 532~nm laser was used to bleach the sapphire surface background in a 14$\times$14 grid pattern with 8~$\mu$m steps. A $\sim$30$\times$ reduction of the background is observed in the low area where the bleaching laser was scanned.}
\end{figure*}

\begin{figure*}
\includegraphics[width=.65\paperwidth]{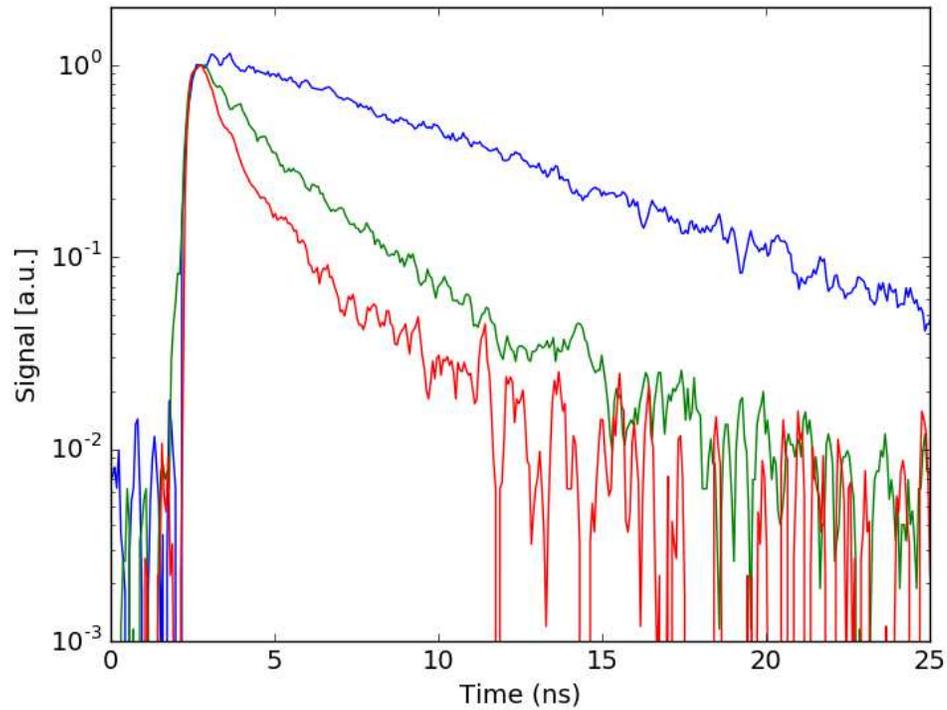}
\caption*{\textbf{Extended Data Fig. 3: Time resolved photon counting of 619~nm fluorescence.} Histograms of the 619~nm fluorescence decay of Ba in SXe (blue), SXe-only (green), and cryoprobe tube (red). The decay lifetime of the Ba fluorescence is 7.0~$\pm$~0.3~ns. The SXe-only and cryoprobe having shorter lifetimes of approximately 3~ns and 1.5~ns respectively.}
\end{figure*}

\end{document}